# Improving speech recognition models with small samples for air traffic control systems

Yi Lin [a], Qin Li [b], Bo Yang [a], Zhen Yan [a], Huachun Tan [c, *], and Zhengmao Chen [a]

[a] *College of Computer Science, Sichuan University, Chengdu 610064, China.*
[b] *School of Mechanical Engineering, Beijing Institute of Technology, Beijing 100081, China.*
[c] *School of Transportation, Southeast University, Nanjing 211189, China.*

ARTICLE INFO

ABSTRACT

In the domain of air traffic control (ATC) systems, efforts to train a practical automatic speech recognition (ASR) model always faces the problem of small training samples since the collection and annotation of speech samples are expert- and domain-dependent task. In this work, a novel training approach based on pretraining and transfer learning is proposed to address this issue, and an improved end-to-end deep learning model is developed to address the specific challenges of ASR in the ATC domain. An unsupervised pretraining strategy is first proposed to learn speech representations from unlabeled samples for a certain dataset. Specifically, a masking strategy is applied to improve the diversity of the sample without losing their general patterns. Subsequently, transfer learning is applied to fine-tune a pretrained or other optimized baseline models to finally achieves the supervised ASR task. By virtue of the common terminology used in the ATC domain, the transfer learning task can be regarded as a sub-domain adaption task, in which the transferred model is optimized using a joint corpus consisting of baseline samples and new transcribed samples from the target dataset. This joint corpus construction strategy enriches the size and diversity of the training samples, which is important for addressing the issue of the small transcribed corpus. In addition, speed perturbation is applied to augment the new transcribed samples to further improve the quality of the speech corpus. Three real ATC datasets are used to validate the proposed ASR model and training strategies. The experimental results demonstrate that the ASR performance is significantly improved on all three datasets, with an absolute character error rate only one-third of that achieved through the supervised training. The applicability of the proposed strategies to other ASR approaches is also validated.

---

* Corresponding author. e-mail: tanhc@seu.edu.cn

# 1. Introduction

As is well known, an air traffic controller (ATCO) guides a flight by sending spoken instructions to the pilot. These instructions contain a wealth of situational context information, embodied in the control intent. In current management systems for air traffic control (ATC), the speech communication between an ATCO and a pilot is a concentrated human-in-the-loop (HITL) [1] procedure. In practice, however, such HITL procedure is considered to present a safety risk and thus to be in need of monitoring using advanced techniques. It is believed that understanding the spoken instructions is an efficient way to monitor the HITL risk and further formulate a closed-loop ATC management system [2]. To this end, automatic speech recognition (ASR) is a powerful interface for human-machine interaction that can allow a machine to automatically understand real-time ATC speech conversations to support further applications. The ASR technique is also expected to be promising for bridging humans (ATCOs and pilots) with the ATC system [3]. Recently, ASR research has attracted significant attention worldwide in many fields. Researchers have applied ASR technique in the ATC domain to address various existing issues, such as operational safety monitoring [4], reducing ATCO workloads [5], and developing simulation interfaces [6].

As a typical supervised learning task, training a practical ASR system strongly depends on the quality of the training corpus [7][8], including the data size, the coverage and diversity of the speech in terms of vocabulary words. With the application of deep learning model in ASR research, the final performance can be greatly improved by updating a large number of trainable parameters of a neural network using data-driven mechanism. To advance the ASR research, various corpora have been built for different common applications and languages, some of which are summarized in Tab. 1. From this table, we can see that it is easy to collect a qualified ASR corpus (up to thousands of hours) to study near-field reading speech recognition for different languages. However, due to the domain-specific characteristics of the civil aviation industry, ATC speech samples are difficult to collect, and there is currently no available speech corpus that is suitable for use in related research. More importantly, the following influencing factors in the ATC domain [9][10] lead to other technical difficulties:

a) Volatile background noise and inferior intelligibility: In ATC scenarios, communication is achieved via radio transmission in the very high frequency (VHF) band, which is always an obstacle to receiving correct and high-quality speech signals. In addition, an ATCO shares the same communication frequency with multiple pilots, resulting in a time-varying system with diverse equipment errors. As a result, the features of ATC speech strongly diverge from those of common speech.

b) Terminology and code-switching: The International Civil Aviation Organization (ICAO) has published the standard procedures for ATC, which specify the only spoken terminology that is allowed to be used during real-time ATC communication. Furthermore, to eliminate misunderstandings caused by homonyms or near-homonyms, some words are given special pronunciations. For example, the English letter 'a' is switched to the pronunciation 'alpha'. Consequently, the annotation of sufficient ATC speech for training a practical ASR system is a highly expert-and domain-dependent, laborious and costly task.

c) Multilingual and accented speech: Although English is the universal language for international flights, in practice, ATCOs are accustomed to communicating with domestic flights in local languages, for example, Chinese is typically used for communicating in mainland China. Additionally, the multilingual speech, such as the use of the English name of the waypoint, will inevitably present a problem for ASR research. Moreover, since pilots come from all over the world, ATC speech is also spoken with different accents. Consequently, it is not possible to apply a common speech corpus to pretrain the ASR model for use in the ATC domain. In other words, only a dedicated corpus and ASR model can meet the requirements of ATC related applications.

**Table 1.** Summary of popular speech corpora for common ASR research.

| No. | Corpus | Language | Domain | Size (hour) | Access |
|---|---|---|---|---|---|
| 1 | LibriSpeech [44] | English | reading novels | 960 | public |
| 2 | TED-LIUM3 [45] | English | TED talks | 452 | public |
| 3 | Switchboard [46] | English | Telephone calls | 260 | public |
| 4 | THCHS30 [47] | Chinese | reading newspapers | 30 | public |
| 5 | AISHELL-V1 [48] | Chinese | multidomain | 500 | public |
| 6 | AISHELL-V2 [49] | Chinese | multidomain | 1000 | application |
| 7 | ATCSpeech [50] | Chinese/English | real ATC | 59 | application |
| 8 | ATCOSIM [51] | English | simulated ATC | 11 | public |
| 9 | LDC94S14 [52] | English | airport | 70 | paid |
| 10 | Airbus [53] | English | pilot | 40 | unavailable |

Although most terms used in the ATC domain are common among different contexts, the vocabulary also presents unique characteristics depending on the flight phase and the location of the control center. For instance, walkie-talkie communication is used for ground (GND) service at an airport, while VHF communication is used for ATC entities, such as the aerodrome control tower (TWR), and the area control center (ACC). Similarly, the vocabulary used for particular waypoints depends on the location, for example, 'PIKAS' is used only by the Chengdu ACC. In summary, it is difficult to build an available speech corpus for the ASR task in the ATC domain. Therefore, training ASR models on small training samples is an essential topic of research in the ATC domain. The small sample problem also arises in many other research fields, such as computer vision (CV) and natural language processing (NLP), in which transfer learning and pretraining are widely used to improve model performance [11][12]. For instance, an optimized classifier for cats can be transferred to train a classifier for dogs in the CV field. Similarly, pretraining can be applied to transfer learned knowledge from one document to another, as done in Bidirectional Encoder Representation from Transformers (BERT) [13], A Lite BERT (ALBERT) [14], etc.

In this work, an improved connectionist temporal classification (CTC) based deep learning ASR model is first proposed to perform the ASR task. The ATC speech is collected with volatile background noise and in multilingual, as a result, the speech features are distributed in a probabilistic space with low cohesion. It is believed that the single-scale convolutional operation may not be able to learn desired representations from the raw speech feature. To address this challenge, a multiscale convolutional neural network (MCNN) architecture is proposed to capture speech features at different scales, enabling the extraction of discriminative and robust speech features from diverse speech.

Subsequently, a combination of training strategy consisting of pretraining and transfer learning is proposed to achieve the desired performance on a small transcribed corpus. Pretraining is performed to learn representative patterns from raw speech data without transcriptions in an unsupervised manner. An improved denoising autoencoder (DAE) with residual connections is developed to perform the unsupervised learning task, in which the neural network is designed to learn both spatial and temporal dependencies. Then, based on the pretrained ASR model and new transcribed samples, a subdomain transfer learning strategy is applied to optimize the ASR model by sharing the model parameters. Based on the fact that most terms are shared among the ATC speech conversations used for different flight phases and by different control centers, the considered transfer learning task is regarded as a subdomain adaption task. In this work, a subdomain can correspond to a certain flight phase (TWR, ACC, etc.) or a location-dependent control center (Beijing, Chengdu, etc.). By virtue of the domain universality, a new ASR model can be the fine-tuned through joint optimization on the transcribed samples used to train the pretrained model and new samples. In our proposal, the unsupervised pretraining strategy allows the ASR model to initially learn the distribution of the speech feature without requiring any labeled samples in the early stage. Subsequently, the ASR task can be completed by means of supervised fine-tuning and transfer learning.

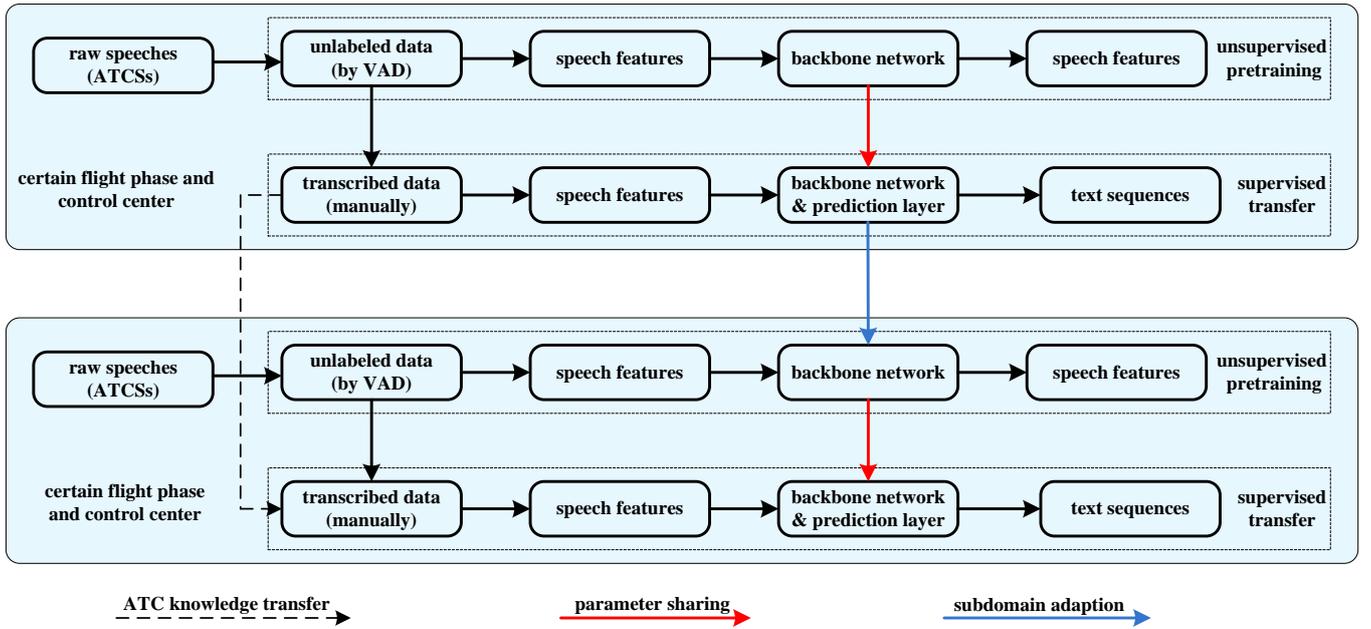

**Fig. 1.** Pipeline for training an ASR model in the ATC domain.

The scheme for training an ASR model in the ATC domain is shown in Fig.1. As illustrated in this figure, the raw speech signals are first segmented into utterances by means of voice activity detection (VAD) to obtain the unlabeled data that are then applied to pretrain the backbone network. After the speech transcriptions are obtained, a prediction layer is appended to the backbone network to achieve the ASR task. Universal ATC knowledge, such as special terminology and call signs, is transferred among different datasets in a process called ATC knowledge transfer. The implementation from pretraining to transfer learning is achieved by sharing the model parameters. Similarly, knowledge of the specific speech characteristics associated with flight phases and control centers (such as waypoints) is learned by transfer learning based on the backbone network in a process called subdomain adaption.

In this study, three real ATC datasets are applied to validate the proposed approach, considering different flight phases and control centers (accents). The experimental results demonstrate that the proposed training strategies are capable of significantly enhancing the ASR performance achieved with small transcribed sample sizes on all three datasets. In summary, this work contributes to ASR research in the ATC domain in the following ways:

1) A MCNN architecture is proposed to serve as the basic for CTC-based speech recognition model, which aims to address the challenges of volatile background noise and multilingual speech by considering the distribution in the feature space at different scales.

2) To solve the small transcribed sample problem for ASR research, a novel pretraining strategy is proposed to learn initial data representations from unlabeled speech in an unsupervised manner, which allows the backbone network of the ASR model to extract high-level speech features to support subsequent supervised ASR optimization.

3) Considering the common terminology used in the ATC domain, a subdomain adaption process is proposed to achieve transfer learning to improve the final ASR performance. The optimized parameters of the baseline model have a powerful ability to extract discriminative features, while the joint corpus construction strategy improves the quality of the training corpus in terms of the data size and diversity.

The rest of this paper is organized as follows. Previous works related to this research are briefly introduced in Section 2. In Section 3, an overview of the deep learning-based ASR model is first present, and the proposed pretraining and transfer learning strategies are also introduced in this section. The experimental configurations are introduced in Section 4, where the experimental results are also discussed. The paper is concluded in Section 5.

**2. Related works**

In deep learning research, a long-standing idea is that the final performance highly depends on the data size, coverage and diversity of the training samples. When faced with the small sample problem, transferring knowledge from another domain is of key importance for advancing the research progress at the beginning of work in a new research field. Two main techniques, i.e., pretraining and transfer learning, have been proposed to address the problem of small sample sizes in many fields. Doersch et al. proposed a self-supervised approach for learning data representations from unlabeled images collected from the Internet using deep learning models [15]. Contrastive predictive coding (CPC) [16] was proposed as the basis for a data-efficient recognition model for CV tasks. Domain transfer learning has been studied to achieve representation learning for document sentiment analysis and image classification [17]. Heterogeneous domain adaptation (HDA) [18] has been studied for the adapt of information across domains with different input feature spaces by means of a learned sparse feature transformation. An online transfer learning framework was studied in [19], in which the cases of both homogeneous and heterogeneous learning were examined. A tensor representation approach has been proposed to achieve domain transfer for CV tasks by aligning the tensor representations from both domains into an invariant tensor subspace [20]. Other transfer learning works have been reviewed in [21].

Currently, many popular and successful pretraining models have been proposed for performing NLP tasks [13][22], such as language generation [23], and language modeling [24]. Word2vec was proposed for extracting high-level representations from word sequences to solve the problem of the sparse encoding (one-hot) of words in NLP research [25]. XLNet [26], a BERT like pretrained model, was implemented based on the autoregressive method using contextualized information. The ALBERT model was proposed to improve the training efficiency [14], while also promoting task performance. The transfer learning technique has also been applied to develop a high-performance dialogue manager by means of domain adaptation [27][28].

Recently, deep learning models have been widely proposed for the ASR task. A deep neural network (DNN) has been proposed to build the data distribution between speech features and text labels [29]. CTC-based approaches, such as the Deepspeech2 (DS2) [30], Jasper [31], CLDNN [32], have achieved excellent ASR performance. Sequence-to-sequence models, such as LAS [33], have also been explored to address the ASR task. In addition, the pretraining and transfer learning techniques have been applied in previous speech processing research. A transfer learning method has been proposed to improve the performance of emotion recognition [34]. Similar research on speaker adaption has been conducted based on mutual information [35]. A CPC-based method has also been studied for learning representations from raw speech [36]. An unsupervised learning method has been proposed for learning data representations (speech features) from one-dimensional speech signal [37][38]. The transfer of speech representations between different languages was studied in [39]. The transformer architecture has been proposed to improve the training efficiency for the ASR task [40]. A multiple self-supervised learning strategy was improved for learning common speech representations [41], in which a single neural encoder is followed by multiple workers that jointly perform different self-supervised tasks. Masked reconstruction has been studied to achieve the unsupervised pretraining of bidirectional speech encoders [42].

## 3. Methodology

In this section, an MCNN based end-to-end ASR model is first proposed to address the specific technical challenges of the ATC domain. In succession, an improved DAE architecture is presented for learning data representations from unlabeled speech via an unsupervised pretraining strategy, which enables the model to obtain a desired parameter optimization to support the final ASR task. Finally, the transfer learning technique is applied in a supervised manner to improve the final ASR performance.

### 3.1. ASR model for the ATC domain

Inspired by state-of-the-art ASR models, a CTC-based end-to-end ASR model is proposed in this work to address the speech recognition issue in the ATC domain. The proposed ASR model is based on a DNN, which is designed on the basis of the MCNN architecture, the long-short term memory (LSTM) architecture, and a fully connected (FC) layer. The overall architecture of the proposed ASR model is illustrated in Fig.2.

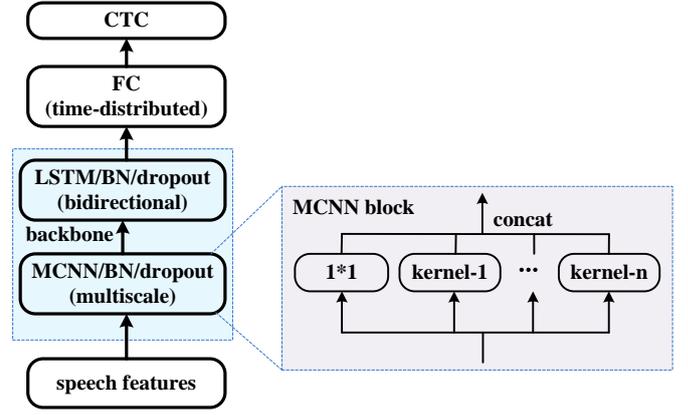

**Fig. 2.** Architecture of the proposed ASR model.

As shown in Fig.2, based on the MCNN architecture, a novel representation learning block is designed to handle the distributed data space caused by the volatile background noise. Two-dimensional convolution operations are applied in the MCNN block, in which convolutional kernels with different local receptive fields capture different data patterns in both the temporal and frequential dimensions at different scales. A 1*1 kernel is used to formulate a residual connection, which is essential for high-efficiency CNN training. A batch normalization (BN) layer and a dropout layer follow each CNN and LSTM layer, with the aims of speeding up the training process and preventing overfitting, respectively. A bidirectional mechanism is implemented in the LSTM layers to extract high-level speech representations by building temporal correlations among the past and future speech frames. The bidirectional long short-term memory (BLSTM) inference rules are formulated as follows, in which $h$ and $b$ denote a hidden unit and the bias, respectively, the notations $\rightarrow$ and $\leftarrow$ represent the inference chains from the past and future directions, respectively, $\Gamma$ denotes the rule of the LSTM cells and $W$ represents vectorized trainable weights.

$$\vec{h}_t = \Gamma(f_t, \vec{h}_{t-1}, \vec{b}), \overleftarrow{h}_t = \Gamma(f_t, \overleftarrow{h}_{t+1}, \overleftarrow{b})$$
$$l_t = W_{\vec{h}l}\vec{h}_t + W_{\overleftarrow{h}l}\overleftarrow{h}_t + b \qquad (1)$$

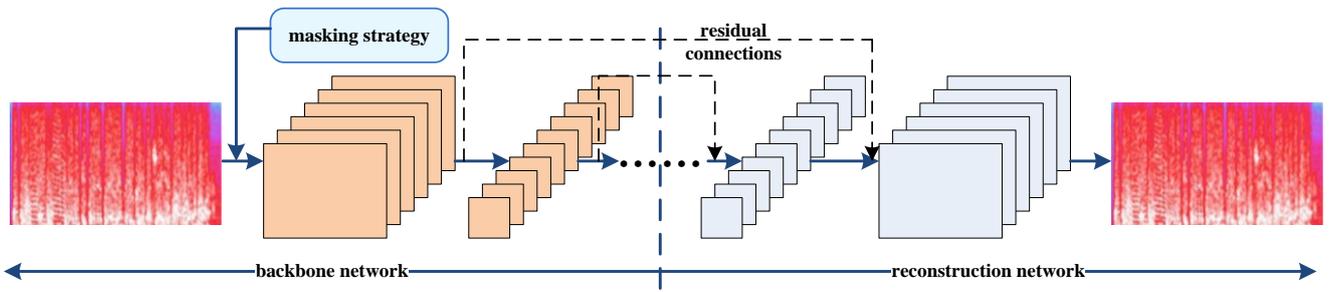

**Fig. 3** Pretraining scheme.

The stacked MCNN/LSTM block (with BN and dropout layers) in the proposed ASR model is defined as the backbone network, where the MCNN and LSTM components serve as the spatial and temporal feature extractors, respectively. Finally, a prediction layer (FC layer) is designed to classify the extracted features as belonging to different modeling units, in a framewise manner. The difference between the predictions and the true labels is evaluated using the CTC loss function, and is then backpropagated to previous layers to support model optimization [43]. In this work, a grapheme-based vocabulary is designed to achieve the multilingual ASR task, in which Chinese characters and English letters are applied to Chinese and English speech, respectively.

As illustrated in Fig.1, the overall training procedure consists of two stages: unsupervised pretraining and supervised transfer learning. In the early stage, since the annotation of ATC speech samples is a laborious task, an unsupervised approach is first applied to pretrain the backbone network. The pretraining process focuses on the mining of universal data patterns from the raw samples, without requiring associated transcriptions. Subsequently, transfer learning for subdomain adaption is performed to optimize the ASR model through supervised learning, with the goal of obtaining a practical ASR system for use in the ATC domain. In the supervised learning task, the network is finally fine-tuned on a joint corpus (consisting of baseline samples and new domain-specific transcribed samples), during which the domain-dependent knowledge is expected to be transferred to support the training of the model.

### 3.2. Unsupervised pretraining with unlabeled speech

To effectively train the backbone network, minimal changes are made for the unsupervised pretraining task. The backbone network can be randomly initialized, or a selected baseline model may be used. The input to the model consists of the 39-dimensional Mel Frequency Cepstrum Coefficient (MFCC) features extracted from the raw speech signals. The output of the model also consists of MFCC feature for the unsupervised learning task.

A schematic illustration of the unsupervised pretraining task is presented in Fig.3. A DAE architecture is designed for the pretraining task, in which the feature extractors are expected to learn characteristic patterns from a certain corpus. In this work, a transposed convolutional network architecture is applied in the reconstruction block, in which residual connections are designed between corresponding layers to improve the trainability. To prevent the shallow copy directly from the input to the output, a masking strategy is applied to generate random noise in the model input, which is beneficial for the learning of robust features. The masking strategy is to randomly modify some parts of the speech features, which forces the model to learn speech representations by optimizing its parameters, instead of simply copying them.

In this work, the masking strategy is dynamic, meaning that new masking patterns are generated each time training samples are fed into the network. This dynamic masking strategy enriches the diversity of the speech features and further enhances the model learning quality, giving the model strong applicability. Once the speech features are fed into the network, 15% of the speech frames

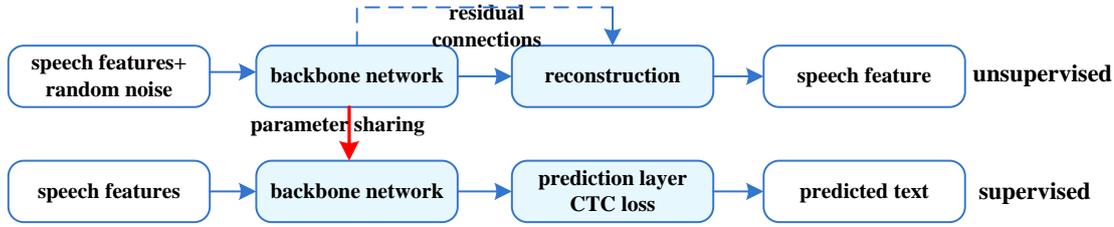

**Fig. 4.** Transfer learning scheme

[40] are randomly selected to be subjected to the masking operation. The features of the selected frames are converted into new features as shown below:

$$\hat{f}_t = \begin{cases} m(f_t) & p < 0.8 \\ \vec{0} & 0.8 \leq p < 0.9 \\ f_t & p \geq 0.9 \end{cases} \quad (2)$$

where $\hat{f}_t$ is the newly generated speech feature based on the raw feature $f_t$, and $t$ is the time index of the speech. The generated speech feature is inferred by means of a piecewise function, where the branch of this function that is ultimately determined based on a random probability $p$. If $p \geq 0.9$, the generated speech feature is identical to the raw one, while if $0.8 \leq p < 0.9$, the new speech feature is set to a zero vector. Otherwise, the case of $p < 0.8$, the masked speech feature is generated in accordance with the following rule:

$$m(f_t) = mean(f_< + f_>) + \xi \quad (3)$$

$$\xi \sim \mathbb{N}(\mu, \delta) \quad (4)$$

where $f_<$ and $f_>$ are the speech features of the nearest left and right neighbors, respectively, of the target frame, and $\xi$ is a random vector drawn from a multiple variable Gaussian distribution whose mean and standard deviation vectors are calculated from the raw inputs. Through the masking operation, the data diversity can be greatly enhanced, while the patterns of the raw data are preserved.

As proven in [33], a down-sampling operation is also applied in the temporal dimension to achieve high-efficiency feature encoding. The mean absolute error (MAE) between the real speech features and the corresponding predicted features (i.e., the output of the reconstruction network) is evaluated to support the optimization of the parameters through the gradient descent operation of the backpropagation algorithm. In the formulas below, $F_i^\xi$ and $\hat{F}_i$ are the noisy input and reconstructed speech features, respectively, corresponding to the $i^{th}$ sample. $N$ is the number of training samples, and $m_i$ is the masking vector, where $T_i$ is the number of speech frames of the $i^{th}$ sample. $m_i^t$ is set to 1 if the $t^{th}$ frame is selected for masking, and to 0 otherwise.

$$\arg\min_\theta \left( \frac{1}{N} \sum_{i=1}^N m_i \left| F_i^\xi - \hat{F}_i \right| \right) \quad (5)$$

$$m_i = \left[ m_i^1, m_i^2, \cdots, m_i^{T_i} \right] \quad (6)$$

### 3.3. Transfer learning for subdomain adaption

With the pretraining strategy and parameter sharing of the transferred model, the backbone network of the ASR model captures both data patterns learned from the corpus of the baseline model and speech representations from a certain unlabeled speech corpus. Consequently, the backbone network has the ability of providing high-level features to support supervised ASR optimization. Subsequently, transfer learning is applied to improve the ASR performance, as shown in Fig.4, which is helpful for solving the small sample problem. The main focuses of the proposed transfer learning technique in this paper are summarized as follows.

1) Supervised learning

Transfer learning is achieved through a fully supervised learning process, in which the whole model is trained on a joint corpus constructed by combining the training corpus of the baseline model with the new transcribed samples from the target dataset.

A time-distributed FC layer is designed to serve as the prediction layer, whose output quantifies the probability of each word based on the speech features. The softmax activation function is used to normalize the output probabilities to ensure that the sum of the probabilities of all words is 1. As shown below, $p_i$ and $\hat{p}_i$ denote the raw neural network output and the normalized probability at the $i^{th}$ index, respectively, and $V$ is the size of the vocabulary. The sum of the probabilities is 1, which indicates that every speech frame $f_t$ will always be predicted to correspond to a word in the vocabulary.

$$\hat{p}_i = \frac{e^{p_i}}{\sum_{i=1}^{V} e^{p_i}} \qquad (7)$$

$$\sum_{i=1}^{V} p(w_i | f_t) = 1, t \in [0,T] \qquad (8)$$

The difference between the predicted and the true labels is evaluated using the CTC loss function to further support parameter optimization. Let the input speech features be $F = \langle f_1, \cdots, f_T \rangle$ and $y_k^t$ denote the probability that the $t^{th}$ frame corresponds to the output label $k$. For a certain input speech, the probability of any output sequence $\pi$ is shown as (9). Therefore, the probability of the final sequence can be obtained by (10), where $\Xi$ is the set of all possible sequences. For example, if '_' is used to denote a blank, then the outputs "a_bb_c" and "_ab_c_" both corresponds to the final output "abc".

$$p(\pi | F) = \prod_{t=1}^{T} y_{\pi_t}^t, \pi_t \in A \qquad (9)$$

$$p(l | F) = \sum_{\pi \in \Xi^{-1}(l)} p(\pi | F) \qquad (10)$$

2) Data augmentation

In this work, due to the special terminology used in the ATC domain, a joint speech corpus is constructed to enhance the diversity of the training samples, i.e., the mapping between the speech features and text labels, while enlarging the data size of the transcribed training corpus used to optimize the model parameters. A joint corpus with higher data diversity is beneficial for overcoming the training challenges presented by the special terminology used in the ATC domain.

In a manner similar to that described above, the speech features of the dataset-dependent vocabulary are provided by the new transcribed training samples, to which a data augmentation strategy is further applied to improve the data diversity. In this work, 50% of the new annotated training samples are randomly selected for speed perturbation by factors of 0.95 and 1.02. To ensure fair comparisons, the raw transcribed samples chosen for augmentation are selected only once and are applied in all of the designed experiments.

3) Training step

In general, the backbone network can be regarded as a feature extractor, while the FC layer serves as a predictor for framewise classification. From the perspective of model training, the backbone network is well optimized by the transferred model and pretraining, while the FC layer is randomly initialized. For the transfer learning technique, the backbone network is first frozen (untrainable) to optimize the FC layer based on the CTC loss. By contrast, the proposed unsupervised pretraining technique relies on reconstruction mechanism based on the DAE architecture (MAE loss). Considering the different loss evaluations for these two steps, it is necessary to unify the target function. Therefore, a new training procedure is proposed for supervised optimization, in which the backbone network is free to be optimized after the 10th training epoch. Thus, the whole model is expected to be fine-tuned to support the ASR task rather than the reconstruction task, which further improves the overall ASR performance.

**Table 2.** Summary of speech corpora used for validation

| Attribute | ZUUU-ACC | ZBAA-GND | ZHCC-TWR | ZPLJ-ACC |
|---|---|---|---|---|
| Control center | Chengdu | Beijing | Zhengzhou | Lijiang |
| Flight phase | ACC | GND | TWR | ACC |
| Accent | Southwest China | North China | Central China | Southwest China |
| Unlabeled data (hour) | - | 145 | 97 | 289 |
| Training set (hour) | 340 | 48 | 47 | 22 |
| Test set (hour) | - | 3.5 | 4 | 2 |
| #Vocabulary words | 1244 | 877 | 1185 | 1052 |
| #New vocabulary | - | 83 | 19 | 54 |

## 4. Experiments and Discussions

### 4.1. Data description

In this work, we mainly focus on the validation of Chinese speech. The speech corpora used in this work are introduced below:

a) Pretraining corpus: Raw speech signals were collected from voice recorder system in the ATCSs, which are segmented into sentence-level utterances, from 2 to 10 seconds in length. This unlabeled corpus collected from certain controlling centers are applied to pretrain the ASR model.

b) Transfer learning corpus: Based on our transcribed samples from the ZUUU-ACC corpus, a baseline model with the proposed architecture was optimized, and the final accuracy was 2.5% according to an evaluation of the character error rate based on the Chinese characters and English letters. The samples used to train the baseline model are applied for transfer learning in combination with new annotated samples (from certain target datasets).

In this work, three real ATC datasets are applied to validate the proposed ASR model and training strategies, considering different flight phases and control centers (accents). The details of the speech corpora are summarized in Tab. 2, where the information for the base-dataset (ZUUU-ACC) is also listed. New vocabulary words appear in each dataset used for subdomain transfer, which is a key factor for transfer learning.

Spectrum examples for each test dataset are shown in Tab.3, where the horizontal and vertical dimensions denote the time and frequency (0-4 kHz), respectively. It can be seen that the speech spectrums from different datasets exhibit unique patterns, which depend on the communication mode and the real-time conditions. Specifically, the speech spectrum in the ZBAA-GND corpus (walkie-talkie) exhibit a particular frequential distribution, whose energy intensity is concentrated at the low frequencies. Even between the VHF datasets, ZHCC-TWR and ZPLJ-ACC, different temporal resolutions are evident in the speech spectrum.

**Table 3.** Examples of speech spectrum from test datasets

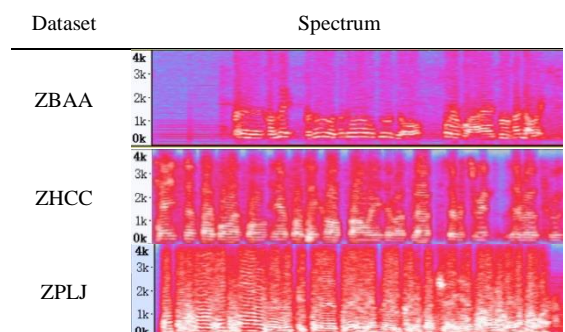

### 4.2. Experimental configurations

The proposed approach is implemented using the Keras framework with a TensorFlow backend. The configuration of the training server is summarized as follows: 2*Intel Core i7-6800K CPUs, 2*NVIDIA GeForce GTX 2080Ti GPUs and 64 GB of memory, running the Ubuntu 16.04 operation system.

The 39-dimensional MFCC features serve as the input to the ASR model, for both unsupervised pretraining and supervised fine-tuning. Considering the feature distribution on the temporal and frequency dimension, two MCNN layers are applied to extract high-level spatial representations from these speech features. Four kernels are designed for each MCNN layer, whose configurations are summarized as follows: 1*1@16, 11*3@32, 13*3@32, and 13*1@16. Here, the notation 1*1@16 indicates that 16 filters with a 1*1 kernel (on temporal and frequency dimension) are applied to learn the data representations. In general, except for the 1*1 kernel, other kernels are designed to deal with the different temporal and frequency resolution. The kernel on the temporal dimension (1, 11 or 13) mainly focuses on the capturing the speech representations with different speech rate, specifically, smaller kernel for higher speech rate. The strides are set to 2 and 1 for the first and second MCNN layers, respectively. The padding mode is set to 'same' to support the MCNN concatenation operation. In succession, five BLSTM layers are designed to mine the temporal dependencies of the speech representations, with each layer containing 512 nodes [4]. The prediction layer is a time-distributed FC layer that classifies the high-level features into vocabulary words.

In this work, the vocabulary contains all words from all four datasets. Moreover, since we mainly focus on the acoustic model, no language model is applied to correct the decoding results from the perspective of semantics in the ATC domain. The Adam optimizer is applied for both pretraining and transfer learning. The initial learning rates are set to $10^{-3}$ and $5*10^{-5}$ for the pretraining and transfer learning, respectively, while the batch sizes are set to 96 and 160 on two GPUs parallel training. To reduce the training loss to a certain level as quickly as possible, all training samples (both unlabeled and labeled) are sorted by their durations in the first training epoch. For the ATC speech, similar durations of different speech samples are expected to indicate that their texts are associated with the same control instructions and share a high similarity. From the second training epoch, all training samples are shuffled to improve the robustness of the ASR model. Moreover, an early stopping strategy based on the validation loss is applied to check the training progress.

In these experiments, the final performance on the ASR task is evaluated in terms of the character error rate (CER) based on Chinese characters and English letters. The CER is calculated by the following rule, where $N$ is the total length of the true label. The notation $I$, $D$, $S$ denote the number of the insertion, deletion and substitution operations, respectively, which are applied to convert the prediction label into the true label.

$$CER = \frac{I+D+S}{N} \times 100\% \quad (11)$$

*4.3. Experiments for validating the proposed ASR model*

For the experiments reported in this section, two baselines are designed to validate the efficiency and effectiveness of the proposed ASR model, namely, DS2 [30] and Jasper 10*3 [31]. In these experiments, all ASR models are trained using a fully supervised learning process, without pretraining or transfer learning. The experimental results in terms of the CER are reported in Tab. 4.

**Table 4.** Results (CER%) for the ASR baselines

| No. | Experiment | ZBAA-GND | ZHCC-TWR | ZPLJ-ACC |
|---|---|---|---|---|
| 1 | DS2 | 17.7 | 10.2 | 13.4 |
| 2 | Jasper | 19.1 | 14.5 | 15.2 |
| 3 | Proposed | 15.9 | 9.7 | 12.8 |

From the results, we can see that the proposed ASR model achieves the best performance among the compared models on all three datasets. Compared to Jasper (a full CNN-based architecture), the recurrent neural network (RNN) based architecture (DS2 and the proposed model) achieve higher accuracy. This can be attributed to the fact that ATC speech is generated in strict compliance with a standard procedure, which allows the RNN block to capture the intrinsic temporal patterns among different words. For the RNN-based model, thanks to the ability of the proposed MCNN block on capturing the diverse speech features, the proposed model achieves higher accuracy than that of the DS2, which also confirms the effectiveness of the technical improvements for the ATC-related corpus. As to the proposed approach, it yields higher performance improvement on the ZBAA-GND dataset than it does on the other two datasets, which can be attributed to the powerful ability of the MCNN architecture to deal with the sparse distribution of the speech features. For different target datasets, the ZHCC-TWR always obtains better accuracy due to higher speech quality and fewer new vocabulary words. On the contrary, the ASR model suffers the largest prediction error on the ZBAA-GND dataset due to inferior speech and more new vocabulary words. In summary, the proposed ASR model can be used to perform the ASR task with considerably high confidence.

*4.4. Experiments for validating pretraining strategy*

In this section, several experiments are conducted to validate the performance of the proposed ASR model with the pretraining strategy, as described below.

a) -P-A: The ASR model is trained only on new transcribed samples from each test dataset, within no pretraining and augmentation. This experiment is also defined as the basic training (baseline) for the validation of the proposed pretraining strategy. For instance, the ASR model for the ZBAA-GND dataset is directly trained on the 48-hour transcribed corpus and evaluated on the 3.5-hour test dataset.

b) -P+A: Following the experiment a), 50% of the transcribed samples are randomly selected for augmentation, and further applied to optimize the ASR model. For the ZBAA-GND dataset, the training corpus for this experiment is summarized as follows: the 48-hour raw transcribed corpus, the 24-hour augmented samples at a *0.95 speech rate, and the 24-hour augmented samples at a *1.02 speech rate. The purpose of this experiment is to prove the effectiveness of the data augmentation strategy.

c) +P-A: The ASR model is pretrained on the unlabeled data and fine-tuned on new transcribed samples from each test dataset. For instance, the ASR model for the ZBAA-GND dataset is first pretrained on the 145-hour unlabeled corpus and then fine-tuned on the 48-hour labeled corpus. Finally, the optimized ASR model is evaluated on the 3.5-hour test dataset. This experiment is designed to confirm the improvement achieved as a result of the pretraining strategy.

d) +P+A: Following experiment c), for each dataset, the new transcribed samples are augmented and used to train the ASR model, as described in b). This experiment is conducted to confirm that both the pretraining and augmentation strategies contribute to enhance the final accuracy.

**Table 5.** Results (CER%) for the pretraining validation

| No. | Experiment | ZBAA-GND | ZHCC-TWR | ZPLJ-ACC |
|---|---|---|---|---|
| 1 | -P-A | 15.9 | 9.7 | 12.8 |
| 2 | -P+A | 12.7 | 7.5 | 9.9 |
| 3 | +P-A | 10.4 | 6.9 | 8.8 |
| 4 | +P+A | 9.2 | 5.0 | 6.8 |

The experimental results are listed in Tab. 5 for all combinations of pretraining and augmentation strategy. It can be seen from these results that both pretraining and augmentation strategies are able to greatly improve the ASR performance. In general, the specific speech representations for a certain dataset are captured by pretraining, while augmentation improves the data size and diversity for the supervised optimization process, in which both of them are to improve the modeling accuracy of the supervised ASR training. As seen from experiments 2 and 3, both pretraining and augmentation are beneficial for improving ASR performance. Compared to experiment 1, an absolute CER reduction of more than 2% is obtained in experiment 2 through augmentation. With the pretraining strategy, the performance promotion is approximately 4% in term of absolute CER reduction for the ZBAA-GND and ZPLJ-ACC datasets, while it is approximately 3% for the ZHCC-TWR dataset. Thus, compared to data augmentation, the pretraining enables greater performance promotion. By combining the two strategies, the desired performance is achieved on all three datasets.

In addition, the ASR performance enhancement further depends on the various attributes of the speech corpus, as summarized below:

a) The speech conditions, concerning the communication mode and the flight phase, have a great influence on the final performance. Since the baseline model was pre-optimized by the speeches of the VHF communication, the ZBAA-GND dataset (walkie-talkie communication) suffers from the largest CER among the three datasets, in which the low-frequency concentrated features are different with that of the VHF communication, and further impact the speech intelligibility. For the speech corpora corresponding to VHF communication, the speech conditions and are similar with that of the baseline model whose data patterns are easy to be captured for model learning. Therefore, the performance improvements on the ZHCC-TWR and ZPLJ-ACC dataset are higher than that of the ZBAA-GND dataset.

b) In addition, the size of the labeled corpus and the vocabulary size also play important roles in achieving desired ASR performance. As demonstrated on the ZBAA-GND dataset, even though the data size is slightly larger than the ZHCC-TWR, the model also fails to achieve comparable performance since there are more new vocabulary. As to the ZPLJ-ACC dataset, the ASR model is able to obtain preferred performance even on only 22-hours labeled corpus since less new words are in the vocabulary. In general, more speech samples with a smaller vocabulary allow the proposed approach to achieve higher performance.

*4.5. Experiments for validating transfer learning strategy*

As mentioned before, a pretrained model was initially built based on the ZUUU-ACC speech corpus, and served as the baseline for transfer learning on the three test datasets. This section reports several experiments conducted to validate the performance of the proposed ASR model with pretraining and transfer learning strategies, as described below:

a) -P-A+T: For each test dataset, the transferred model is directly trained on a joint corpus constructed by combining the base dataset with new transcribed samples, with no pretraining and augmentation. For instance, the baseline ASR model for ZBAA-GND is optimized on the 340-hour base data and 48-hour speech corpus and evaluated on the 3.5-hour test set. With this experiment, we mainly focus on the influence of transfer learning on the ASR accuracy.

b) -P+A+T: Following experiment a), 50% of the newly transcribed samples are randomly selected for augmentation. Accordingly, a speech corpus with a total duration of 436-hour (340+48+48) is applied to optimize the baseline ASR model for the ZBAA-GND dataset. The purpose of this experiment is to joint validate the data augmentation and the pretraining strategies.

c) +P-A+T: For each dataset, the ASR model is first pretrained on the unlabeled data and then fine-tuned on a joint corpus consisting of the base dataset and new transcribed samples. For instance, for the ZBAA-GND dataset, the backbone network of the baseline ASR model is pretrained on the 145-hour unlabeled speech dataset and fine-tuned on the 388-hour joint corpus (340+48), and it is finally evaluated on the 3.5-hour test dataset. This experiment is designed to validate the proposed pretraining and transfer learning strategies.

d) +P+A+T: Following experiment c), the augmented corpus is also applied to fine-tune the baseline ASR model for each dataset. In this experiment, all three proposed strategies are applied to solve the problem of small samples for ASR research in the ATC domain.

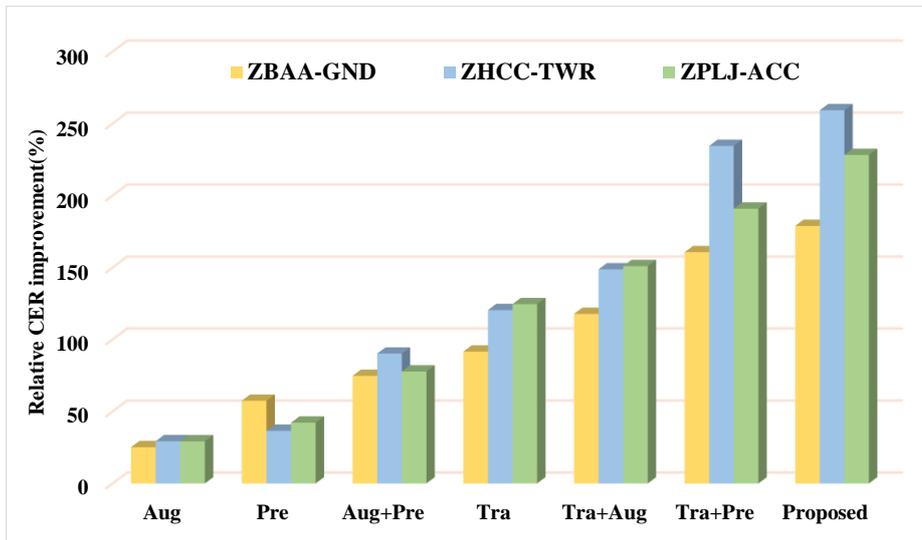

**Fig. 5.** Relative CER improvements obtained with different strategies on the three datasets.

**Table 6.** Experimental results (CER%) for transfer learning validation

| No. | Experiments | ZBAA-GND | ZHCC-TWR | ZPLJ-ACC |
|---|---|---|---|---|
| 0 | -P-A-T | 15.9 | 9.7 | 12.8 |
| 1 | -P-A+T | 8.3 | 4.4 | 5.7 |
| 2 | -P+A+T | 7.3 | 3.9 | 5.1 |
| 3 | +P-A+T | 6.0 | 2.8 | 4.5 |
| 4 | +P+A+T | 5.5 | 2.7 | 3.8 |

In Tab.6, experiment 0 corresponds to the baseline model with the supervised training. From the results of experiments 1-4, we can see that the transfer learning technique (sub-domain adaption) significantly enhances the ASR performance, which is even higher than that achieved with the joint pretraining and augmentation strategy (experiments 0 and 1). By combining the pretraining, transfer learning and augmentation strategies, the desired performance improvement is obtained with small transcribed sample sizes for all three datasets, with CERs that are only approximately one-third of those achieved with the full supervised training approach. The relative CER improvements achieved with the different strategies on all three datasets are shown in Fig.5, where 'Aug', 'Pre' and 'Tra' denote augmentation, pretraining and transfer learning, respectively. As shown in the figure, all the proposed three techniques are beneficial to improve the final performance, in which their combinations are able to contributes more performance improvements. We attribute these improvements to the following two factors:

1) The baseline model for knowledge transfer lays a solid foundation for supporting the supervised training of the ASR model on the new dataset. The parameters of the baseline model have already been optimized on another ATC-related speech corpus, enabling the extraction of discriminative features for the word classification. The training on the new target dataset is mainly to build the mappings between speech frames and new vocabulary words, and fine-tune the optimized mappings based on the target speech conditions.

2) The size and diversity of the training corpus are highly enhanced by the base dataset. For the common ATC terminology, both the base data and the new transcribed samples provide information on the mapping between the speech frames and the text labels to support the optimization of the ASR model, thus, greatly improving the data diversity for the transfer of ATC knowledge.

Specifically, as demonstrated by the results for the ZHCC-TWR dataset, since the terms used by the control tower are almost entirely common ones for different locations (there are only 19 new vocabulary words in this corpus), the performance improvement is prominent and the final CER is only 2.7%. This represents an excellent ASR performance on only 47-hour transcribed samples for air traffic-related research and further validates the strategies proposed in this work. For the ZPLJ-ACC dataset, although the flight phase is the same as that of the baseline model (ZUUU-ACC), the improvement (3.8% CER) is not comparable to that achieved on the ZHCC-TWR dataset due to the smaller sample size and the larger number of new vocabulary words. In regard to the ZBAA-GND dataset, the performance enhancement is also affected by the domain diversity and speech quality, which are inferior to those in the other two datasets. In short, similar speech conditions with less new vocabulary words is able to yield the higher performance improvement by the proposed transfer learning strategy. Nonetheless, the subdomain adaption strategy is always beneficial for improving the ASR performance on a new corpus in the ATC domain.

As seen from the results, the data augmentation strategy offers only minor contribution if transfer learning is already being applied to train the ASR model, corresponding to an absolute CER reduction of approximately 0.5% for all three datasets. This is because the data diversity is already enhanced by the inclusion of the corpus used to train the baseline model, especially for the same flight phase with the base corpus, resulting in only marginal improvement with the augmentation. As a comparison, the pretraining is more important augmentation for improving the model performance (about 1.5% CER reduction) since it focuses on learning the intrinsic data patterns to support the supervised optimization on the target corpus. In combination with transfer learning, the pretraining strategy leads to an absolute CER reduction of approximately 2% for all datasets, thus, validating the effectiveness of the pretraining and transfer learning strategies in this work. Finally, when all three strategies are combined, the ASR model achieves the desired performance with only a small labeled speech corpus, which is the ultimate goal in this work.

*4.6. Experiments for evaluating applicability*

In this section, two experiments are further designed to prove the applicability of the proposed approach, in which the pretraining and transfer learning strategies are applied to train two different baseline models, DS2 and Jasper 10*3. The experimental results are reported in Tab.7.

**Table 7.** Experimental results (CER%) for applicability evaluation

| No. | Model | Training | ZBAA-GND | ZHCC-TWR | ZPLJ-ACC |
|---|---|---|---|---|---|
| 1 | DS2 | supervised | 17.7 | 10.2 | 13.4 |
| 2 | | proposed | 6.1 | 4.0 | 6.1 |
| 3 | Jasper | supervised | 19.1 | 14.5 | 15.2 |
| 4 | | proposed | 7.3 | 4.8 | 5.5 |

It can be seen from these results that although the proposed training strategies are developed for the MCNN based ASR model, they also work on other baseline models. Both the DS2 and Jasper models show the performance improvements with the application of the proposed pretraining and transfer learning strategy. Compared to results of the supervised training, the experimental results exhibit the same trend on different datasets, i.e., performance improvement from ZHCC-TWR, ZPLJ-ACC and the ZBAA-GND. The experimental results also prove the capacity of different models for ASR research in the ATC domain, including the data size and diversity, and the new vocabulary.

In addition, with the same training strategies, the proposed ASR model achieves higher performance compared to the comparative approaches on all the test datasets. Considering the experimental configurations (the same dataset and training parameters), it can be concluded that the proposed ASR model with the MCNN block makes great contributions to improve the final performance since it is designed to address the certain technical difficulties for the ASR research in the ATC domain.

**5. Conclusions**

In this paper, a new training approach is proposed to achieve ASR task with small transcribed speech samples in the air traffic domain. The proposed approach consists of two stages:

unsupervised pretraining and supervised transfer learning. The proposed pretraining procedure is applied to learn speech representations from unlabeled speech samples, in this procedure, a masking strategy is applied to ensure efficient training. The transfer learning task considered in this work is regarded as a subdomain adaption task since the ATC speech vocabularies for different flight phases and control centers largely share common terminology. The speed perturbation is also performed to improve the size and diversity of the labeled training corpus. An improved deep learning-based end-to-end model is developed for the ASR task, where the model architecture is based on MCNN, LSTM, and CTC blocks. To handle the distributed data patterns of ATC speech, an MCNN block is designed to capture speech representations at different scales. In the transfer learning stage, the baseline ASR model is further fine-tuned on a joint corpus constructed by combining the samples used to train the baseline model with new transcribed samples from a target dataset. This joint corpus construction strategy enhances the size and diversity of the training samples used to transfer the learned ATC knowledge to a new subdomain. Three real ATC datasets are applied to validate the proposed training strategy, considering different flight phases and control centers (accents). The experimental results show that the proposed training approach is able to significantly improve the ASR performance compared with that achieved through supervised training. The characteristics of ATC speech are successfully captured by pretraining with the backbone network. In addition, it is believed that transfer learning based on a joint corpus is particularly important for training a practical ASR system for use in the ATC domain.

In the future, we plan to further apply the adversarial learning mechanism to perform online data augmentation and formulate a unified framework with the ASR model.

**Acknowledgments**

This work was jointly supported by the National Natural Science Foundation of China (Grant No: 62001315, 61620106002), and the Sichuan Provincial S&T Projects (Grant No. 2020YFG0327)

**References and notes**

[1] F.H. Hawkins, Human factors in flight, Appl. Ergon. 19 (1988) 337. doi:10.1016/0003-6870(88)90093-2.
[2] H.D. Kopald, A. Chanen, S. Chen, E.C. Smith, R.M. Tarakan, Applying automatic speech recognition technology to Air Traffic Management, in: 2013 IEEE/AIAA 32nd Digit. Avion. Syst. Conf., IEEE, 2013: pp. 6C3-1-6C3-15. doi:10.1109/DASC.2013.6712620.
[3] R. de Cordoba, J. Ferreiros, R. San-Segundo, J. Macias-Guarasa, J.M. Montero, F. Fernandez, L.F. D'Haro, J.M. Pardo, Air traffic control speech recognition system cross-task and speaker adaptation, IEEE Aerosp. Electron. Syst. Mag. 21 (2006) 12–17. doi:10.1109/MAES.2006.1705165.
[4] Y. Lin, L. Deng, Z. Chen, X. Wu, J. Zhang, B. Yang, A Real-Time ATC Safety Monitoring Framework Using a Deep Learning Approach, IEEE Trans. Intell. Transp. Syst. 21 (2020) 4572–4581. doi:10.1109/TITS.2019.2940992.
[5] H. Helmke, O. Ohneiser, T. Muhlhausen, M. Wies, Reducing controller workload with automatic speech recognition, in: 2016 IEEE/AIAA 35th Digit. Avion. Syst. Conf., IEEE, 2016: pp. 1–10. doi:10.1109/DASC.2016.7778024.
[6] H.-L. Lu, V.H. Cheng, D. Ballinger, A. Fong, J. Nguyen, S. Jones, S.E. Cowart, A Speech-Enabled Simulation Interface Agent for Airspace System Assessments, in: AIAA Model. Simul. Technol. Conf., American Institute of Aeronautics and Astronautics, Reston, Virginia, 2015. doi:10.2514/6.2015-0148.
[7] Y. Lin, L. Li, H. Jing, B. Ran, D. Sun, Automated traffic incident detection with a smaller dataset based on generative adversarial networks, Accid. Anal. Prev. 144 (2020) 105628. doi:10.1016/j.aap.2020.105628.
[8] Y. Lin, J. Zhang, H. Liu, Deep learning based short-term air traffic flow prediction considering temporal–spatial correlation, Aerosp. Sci. Technol. 93 (2019) 105113. doi:10.1016/j.ast.2019.04.021.
[9] Y. Lin, X. Tan, B. Yang, K. Yang, J. Zhang, J. Yu, Real-time Controlling Dynamics Sensing in Air Traffic System, Sensors. 19 (2019) 679. doi:10.3390/s19030679.
[10] Y. Lin, D. Guo, J. Zhang, Z. Chen, B. Yang, A Unified Framework for Multilingual Speech Recognition in Air Traffic Control Systems, IEEE Trans. Neural Networks Learn. Syst. (2020) 1–13. doi:10.1109/TNNLS.2020.3015830.
[11] J. Yosinski, J. Clune, Y. Bengio, H. Lipson, How transferable are features in deep neural networks?, Adv. Neural Inf. Process. Syst. (2014). http://arxiv.org/abs/1411.1792.
[12] M. Long, Y. Cao, J. Wang, M.I. Jordan, Learning Transferable Features with Deep Adaptation Networks, 32nd Int. Conf. Mach. Learn. ICML 2015. (2015). http://arxiv.org/abs/1502.02791.
[13] J. Devlin, M.-W. Chang, K. Lee, K. Toutanova, BERT: Pre-training of Deep Bidirectional Transformers for Language Understanding, (2018). http://arxiv.org/abs/1810.04805.
[14] Z. Lan, M. Chen, S. Goodman, K. Gimpel, P. Sharma, R. Soricut, ALBERT: A Lite BERT for Self-supervised Learning of Language Representations, (2019). http://arxiv.org/abs/1909.11942.
[15] C. Doersch, A. Gupta, A.A. Efros, Unsupervised visual representation learning by context prediction, in: Proc. IEEE Int. Conf. Comput. Vis., 2015. doi:10.1109/ICCV.2015.167.
[16] O.J. Hénaff, A. Srinivas, J. De Fauw, A. Razavi, C. Doersch, S.M.A. Eslami, A. van den Oord, Data-Efficient Image Recognition with Contrastive Predictive Coding, (2019). http://arxiv.org/abs/1905.09272.
[17] Y. Ganin, E. Ustinova, H. Ajakan, P. Germain, H. Larochelle, F. Laviolette, M. Marchand, V. Lempitsky, Domain-adversarial training of neural networks, in: Adv. Comput. Vis. Pattern Recognit., 2017. doi:10.1007/978-3-319-58347-1_10.
[18] C. Shen, Y. Guo, J. Zhu, I. Takeuchi, Unsupervised Heterogeneous Domain Adaptation with Sparse Feature Transformation, 2018.

[19] P. Zhao, S.C.H. Hoi, OTL: A framework of online transfer learning, in: ICML 2010 - Proceedings, 27th Int. Conf. Mach. Learn., 2010.

[20] H. Lu, L. Zhang, Z. Cao, W. Wei, K. Xian, C. Shen, A. Van Den Hengel, When Unsupervised Domain Adaptation Meets Tensor Representations, in: Proc. IEEE Int. Conf. Comput. Vis., 2017. doi:10.1109/ICCV.2017.72.

[21] S.J. Pan, Q. Yang, A Survey on Transfer Learning, IEEE Trans. Knowl. Data Eng. 22 (2010) 1345–1359. doi:10.1109/TKDE.2009.191.

[22] A. Baevski, S. Edunov, Y. Liu, L. Zettlemoyer, M. Auli, Cloze-driven Pretraining of Self-attention Networks, (2019). http://arxiv.org/abs/1903.00778.

[23] S. Edunov, A. Baevski, M. Auli, Pre-trained Language Model Representations for Language Generation, (2019). doi:10.18653/v1/n19-1409.

[24] G. Lample, A. Conneau, Cross-lingual Language Model Pretraining, (2019). http://arxiv.org/abs/1901.07291.

[25] T. Mikolov, K. Chen, G. Corrado, J. Dean, Efficient Estimation of Word Representations in Vector Space, 1st Int. Conf. Learn. Represent. ICLR 2013 - Work. Track Proc. (2013). http://arxiv.org/abs/1301.3781.

[26] Z. Yang, Z. Dai, Y. Yang, J. Carbonell, R. Salakhutdinov, Q. V. Le, XLNet: Generalized Autoregressive Pretraining for Language Understanding, (2019). http://arxiv.org/abs/1906.08237.

[27] M. Gasic, N. Mrksic, L.M. Rojas-Barahona, P.-H. Su, S. Ultes, D. Vandyke, T.-H. Wen, S. Young, Dialogue manager domain adaptation using Gaussian process reinforcement learning, (2016). http://arxiv.org/abs/1609.02846.

[28] V.-K. Tran, L.-M. Nguyen, Adversarial Domain Adaptation for Variational Neural Language Generation in Dialogue Systems, (2018). http://arxiv.org/abs/1808.02586.

[29] A. Abe, K. Yamamoto, S. Nakagawa, Robust speech recognition using DNN-HMM acoustic model combining noise-aware training with spectral subtraction, in: Proc. Annu. Conf. Int. Speech Commun. Assoc. INTERSPEECH, 2015: pp. 2849–2853.

[30] D. Amodei, R. Anubhai, E. Battenberg, C. Case, J. Casper, B. Catanzaro, J. Chen, M. Chrzanowski, A. Coates, G. Diamos, E. Elsen, J. Engel, L. Fan, C. Fougner, T. Han, A. Hannun, B. Jun, P. LeGresley, L. Lin, N. Narang, A. Ng, S. Ozair, R. Prenger, J. Raiman, S. Satheesh, D. Seetapun, S. Sengupta, Y. Wang, Z. Wang, C. Wang, B. Xiao, D. Yogatama, J. Zhan, Z. Zhu, Deep Speech 2: End-to-End Speech Recognition in English and Mandarin, 33rd Int. Conf. Mach. Learn. ICML 2016. (2015). http://arxiv.org/abs/1512.02595.

[31] J. Li, V. Lavrukhin, B. Ginsburg, R. Leary, O. Kuchaiev, J.M. Cohen, H. Nguyen, R.T. Gadde, Jasper: An End-to-End Convolutional Neural Acoustic Model, in: Interspeech 2019, ISCA, ISCA, 2019: pp. 71–75. doi:10.21437/Interspeech.2019-1819.

[32] T.N. Sainath, O. Vinyals, A. Senior, H. Sak, Convolutional, Long Short-Term Memory, fully connected Deep Neural Networks, in: 2015 IEEE Int. Conf. Acoust. Speech Signal Process., IEEE, 2015: pp. 4580–4584. doi:10.1109/ICASSP.2015.7178838.

[33] W. Chan, N. Jaitly, Q. Le, O. Vinyals, Listen, attend and spell: A neural network for large vocabulary conversational speech recognition, in: 2016 IEEE Int. Conf. Acoust. Speech Signal Process., IEEE, 2016: pp. 4960–4964. doi:10.1109/ICASSP.2016.7472621.

[34] Z. Lian, Y. Li, J. Tao, J. Huang, Improving speech emotion recognition via Transformer-based Predictive Coding through transfer learning, (2018). http://arxiv.org/abs/1811.07691.

[35] M. Ravanelli, Y. Bengio, Learning Speaker Representations with Mutual Information, (2018). http://arxiv.org/abs/1812.00271.

[36] A. van den Oord, Y. Li, O. Vinyals, Representation Learning with Contrastive Predictive Coding, (2018). http://arxiv.org/abs/1807.03748.

[37] S. Schneider, A. Baevski, R. Collobert, M. Auli, wav2vec: Unsupervised Pre-Training for Speech Recognition, in: Interspeech 2019, ISCA, ISCA, 2019: pp. 3465–3469. doi:10.21437/Interspeech.2019-1873.

[38] A. Baevski, S. Schneider, M. Auli, vq-wav2vec: Self-Supervised Learning of Discrete Speech Representations, ArXiv. (2019). doi:1910.05453v3.

[39] J. Kunze, L. Kirsch, I. Kurenkov, A. Krug, J. Johannsmeier, S. Stober, Transfer Learning for Speech Recognition on a Budget, in: Proc. 2nd Work. Represent. Learn. NLP, Association for Computational Linguistics, Stroudsburg, PA, USA, 2017: pp. 168–177. doi:10.18653/v1/W17-2620.

[40] D. Jiang, X. Lei, W. Li, N. Luo, Y. Hu, W. Zou, X. Li, Improving Transformer-based Speech Recognition Using Unsupervised Pre-training, (2019). http://arxiv.org/abs/1910.09932.

[41] S. Pascual, M. Ravanelli, J. Serrà, A. Bonafonte, Y. Bengio, Learning Problem-Agnostic Speech Representations from Multiple Self-Supervised Tasks, in: Interspeech 2019, ISCA, ISCA, 2019: pp. 161–165. doi:10.21437/Interspeech.2019-2605.

[42] W. Wang, Q. Tang, K. Livescu, Unsupervised Pre-Training of Bidirectional Speech Encoders via Masked Reconstruction, in: 2020. doi:10.1109/icassp40776.2020.9053541.

[43] Y. LeCun, B. Boser, J.S. Denker, D. Henderson, R.E. Howard, W. Hubbard, L.D. Jackel, Backpropagation Applied to Handwritten Zip Code Recognition, Neural Comput. 1 (1989) 541–551. doi:10.1162/neco.1989.1.4.541.

[44] V. Panayotov, G. Chen, D. Povey, S. Khudanpur, Librispeech: An ASR corpus based on public domain audio books, in: 2015 IEEE Int. Conf. Acoust. Speech Signal Process., IEEE, 2015: pp. 5206–5210. doi:10.1109/ICASSP.2015.7178964.

[45] F. Hernandez, V. Nguyen, S. Ghannay, N. Tomashenko, Y. Estève, TED-LIUM 3: Twice as Much Data and Corpus Repartition for Experiments on Speaker Adaptation, in: 2018: pp. 198–208. doi:10.1007/978-3-319-99579-3_21.

[46] J. Godfrey, E. Holliman, https://catalog.ldc.upenn.edu/LDC97S62, Linguist. Data Consort. (1997).

[47] D. Wang, X. Zhang, THCHS-30 : A Free Chinese Speech Corpus, ArXiv:1512.01882. (2015). http://arxiv.org/abs/1512.01882.

[48] H. Bu, J. Du, X. Na, B. Wu, H. Zheng, AISHELL-1: An open-source Mandarin speech corpus and a speech recognition baseline, in: 2017 20th Conf. Orient. Chapter Int. Coord. Comm. Speech Databases Speech I/O Syst. Assess., IEEE, 2017: pp. 1–5. doi:10.1109/ICSDA.2017.8384449.

[49] J. Du, X. Na, X. Liu, H. Bu, AISHELL-2: Transforming Mandarin ASR Research Into Industrial Scale, ArXiv Prepr. ArXiv1808.10583. (2018). http://arxiv.org/abs/1808.10583.

[50] B. Yang, X. Tan, Z. Chen, B. Wang, M. Ruan, D. Li, Z. Yang, X. Wu, Y. Lin, ATCSpeech: A Multilingual Pilot-Controller Speech Corpus from Real Air Traffic Control Environment, in: Interspeech 2020, ISCA, ISCA, 2020: pp. 399–403. doi:10.21437/Interspeech.2020-1020.

[51] K. Hofbauer, S. Petrik, H. Hering, The ATCOSIM corpus of non-prompted clean air traffic control speech, in: Proc. 6th Int. Conf. Lang. Resour. Eval. Lr. 2008, 2008.

[52] J. Godfrey, https://catalog.ldc.upenn.edu/LDC94S14A, Linguist. Data Consort. (1994).

[53] T. Pellegrini, J. Farinas, E. Delpech, F. Lancelot, The Airbus Air Traffic Control Speech Recognition 2018 Challenge: Towards ATC Automatic Transcription and Call Sign Detection, in: Interspeech 2019, ISCA, ISCA, 2019: pp. 2993–2997. doi:10.21437/Interspeech.2019-1962.